\documentclass[fleqn,usenatbib]{mnras}

\usepackage{newtxtext,newtxmath}

\usepackage[T1]{fontenc}

\DeclareRobustCommand{\VAN}[3]{#2}
\let\VANthebibliography\thebibliography
\def\thebibliography{\DeclareRobustCommand{\VAN}[3]{##3}\VANthebibliography}

\usepackage{graphicx}	
\usepackage{amsmath}	

\title[Cometary ion drift energy and temperature]{Cometary ion drift energy and temperature at comet 67P-Churyumov/Gerasimeko}

\author[H. N. Williamson et al.]{
Hayley N. Williamson,$^{1}$\thanks{E-mail: hayley.williamson@irf.se}
Annie Johansson,$^{2}$
Romain Canu-Blot,$^{1,3}$
Gabriella Stenberg Wieser,$^{1,3}$ 
\newauthor
Hans Nilsson,$^{1,3}$
Fredrik L. Johansson,$^{4}$
and Anja Moeslinger $^{1,3}$
\\
$^{1}$Swedish Institute of Space Physics, Kiruna 981 92, Sweden\\
$^{2}$Lule\aa \ University of Technology\\
$^{3}$Ume\aa \ University\\
$^{4}$ESA/ESTEC, Noordwijk, Netherlands
}

\date{Accepted XXX. Received YYY; in original form ZZZ}

\pubyear{2024}

\begin{document}
\label{firstpage}
\pagerange{\pageref{firstpage}--\pageref{lastpage}}
\maketitle

\begin{abstract}
The Ion Composition Analyzer (ICA) on the Rosetta spacecraft observed both the solar wind and the cometary ionosphere around comet 67P/Churyumov-Gerasimenko for nearly two years. However, observations of low energy cometary ions were affected by a highly negative spacecraft potential, and the ICA ion density estimates were often much lower than plasma densities found by other instruments. Since the low energy cometary ions are often the highest density population in the plasma environment, it is nonetheless desirable to understand their properties. To do so, we select ICA data with densities comparable to those of Rosetta's Langmuir Probe (LAP)/Mutual Impedance Probe throughout the mission. We then correct the cometary ion energy distribution of each energy-angle scan for spacecraft potential and fit a drifting Maxwell-Boltzmann distribution, which gives an estimate of the drift energy and temperature for 3521 scans. The resulting drift energy is generally between 11--18 eV and the temperature between 0.5--1 eV. The drift energy shows good agreement with published ion flow speeds from LAP during the same time period and is much higher than the cometary neutral speed. We see additional higher energy cometary ions in the spectra closest to perihelion, which can either be a second Maxwellian or a kappa distribution. The energy and temperature are negatively correlated with heliocentric distance, but the slope of the change is small. It cannot be quantitatively determined whether this trend is primarily due to heliocentric distance or spacecraft distance to the comet, which increased with decreasing heliocentric distance.
\end{abstract}




\section{Introduction} \label{sec:Intro}

The interaction of the solar wind with a comet has been observed at multiple comets, primarily flybys of comets 21P/Giacobini-Zimmer \citep{Brandt1985}, 1P/Halley \citep{Gringauz1986}, and 26P/Grigg-Skjellerup \citep{Grensemann1993}. However, a better understanding of comet magnetospheres required long-term observations. Subsequently, the Rosetta mission, designed to rendezvous with and orbit a comet, was launched in 2004. It arrived at comet 67P/Churyumov-Gerasimenko on August 8, 2014 and orbited until September 30, 2016, when it was deorbited into the comet nucleus. While comet 67P was less active than the comets of previous flyby missions, Rosetta's escort of the comet from 3.6 AU to perihelion at 1.2 AU and back out to 3.8 AU offered the first opportunity to observe a changing comet. In particular, the plasma environment of comet 67P changed drastically with heliocentric distance, beginning with a solar wind-dominated environment that became increasingly cometary ion-dominated as higher insolation led to higher outgassing rates \citep{Nilsson2015a,Nilsson2017}. Neutral particles outgassed from the nucleus are subsequently ionized by photoionization and electron impact ionization. These newborn cometary ions radially expand from the nucleus and are accelerated by electric fields \citep{Nilsson2018,moeslinger_indirect_2023}, to eventually be picked up by and mass load the solar wind \citep[e.g.,][]{Coates2004,Szego2000}. 

The plasma environment around the comet is largely dominated by three different types of electric fields, all of which vary with both distance from the comet and comet distance from the sun. When ions are born from the neutral cometary atmosphere, they originally expand radially away from the nucleus. The electrons have a much higher thermal speed than the ions, creating a pressure gradient that leads to a charge separation and hence an ambipolar electric field. This ambipolar electric field accelerates the ions such that their flow velocity is significantly greater than that of the neutrals \citep{vigren_1d_2017} and is strongest close to the comet nucleus \citep{Deca2019}. The ambipolar field magnitude is proportional to the electron number density gradient, $\sim 1/r$ \citep{vigren_1d_2017}. Thus, ions are continuously accelerated as they travel radially outwards from the comet. Higher energy ions are therefore assumed to have been born further away from the spacecraft than lower energy ions.

However, the ambipolar field is not the only electric field the cometary ions experience. Because of the small size of the comet ionosphere and weak interplanetary magnetic field strength, the ion gyroradius (particularly for the cometary ions) is much larger than the ionosphere size, even close to perihelion, making them essentially unmagnetized \citep{williamson_development_2022}. However, electrons have a much smaller gyroradius and as such experience $\vec{E} \times \vec{B}$ drift. The different motions of the ions and electrons thus create a polarisation electric field in an anti-sunward direction. The polarisation electric field is strongest at intermediate heliocentric distances: low column densities far from perihelion will cause a small electric field magnitude, while at high column densities near perihelion, the cometary ionosphere modifies the solar wind electric field significantly enough that the ionosphere does not experience the tailward electric field \citep{Nilsson2018}. Eventually, at the furthest distances from the nucleus, the ions are picked up by the solar wind convective electric field and begin to gyrate in the interplanetary magnetic field. Both low energy newborn ions, in the process of being accelerated by the ambipolar electric field, and higher energy pickup ions were observed by Rosetta at comet 67P \citep{Bercic2018,Masunaga2019,Nicolaou2017,Nilsson2017,StenbergWieser2017}.

While Rosetta orbited relatively close to the comet nucleus, the expansion of the comet magnetosphere with decreasing heliocentric distance allowed for measurements of various solar wind-cometary ion interaction regions \citep{Williamson2020}. Near perihelion, the cometary magnetosphere was extensive enough to exclude the solar wind entirely, called the solar wind ion cavity, and at times even exclude the interplanetary magnetic field, forming a diamagnetic cavity \citep{Goetz2016a,Goetz2016}. The time period where the cometary ionosphere reaches a comparable pressure to the solar wind and as such is able to significantly deflect the solar wind is of particular interest, as it indicates a fully-formed induced magnetosphere. Observations just outside of the solar wind ion cavity, where cometary ion momentum flux is greater than the solar wind momentum flux, are analogous to being just inside a magnetopause. Solar wind ion distributions during this time are similar to those found in a magnetosheath, suggesting the presence of a bow shock \citep{williamson_development_2022}. This highly dynamic region is likely to have formed various types of plasma instabilities (e.g. Kelvin-Helmholtz instabilities seen at Venus and Mars, \citet{poh_growth_2021,Ruhunusiri2016,Wolff1980}); however, thorough knowledge of the plasma pressure tensor is useful for analysis of such instabilities. Thus far, such analysis has been complicated by incomplete knowledge of ion temperatures at comet 67P due to instrumental effects, as described below.

Rosetta routinely experienced highly negative spacecraft potential ($V_{s/c}$), sometimes exceeding -15 V \citep{Odelstad2015b,Odelstad2017}, a result of exposed positively charged connectors on the large solar panels \citep{johansson_charging_2020}. The negative spacecraft charging has the effect of repelling electrons and attracting lower energy positive ions. This affects plasma measurements, such as those made by the Ion Composition Analyzer (ICA) \citep{Nilsson2007}, part of the Rosetta Plasma Consortium \citep{Carr2007}. ICA was designed to study the interaction between the solar wind and cometary ions by measuring positive ions through a wide range of energies and masses. As such, it is capable of measuring the low energy ions that contribute the bulk of the pressure close to perihelion. However, the spacecraft potential both accelerates and distorts the trajectories of these ions, changing their energy distribution. This has the largest effect on ions with energies $ \lessapprox |2 V_{s/c}|$ \citep{Bergman2020,Bergman2021,Bergman2021a}. In addition, ion densities measured by ICA were often much lower than those measured by the RPC Langmuir Probe (LAP) \citep{Eriksson2007} and Mutual Impedance Probe (MIP) \citep{trotignon_rpc-mip_2007}. It is unclear why this is the case, as the ICA densities can be up to two orders of magnitude lower than those found by LAP/MIP. One likely possibility is that the distortion of low energy ion trajectories by the spacecraft potential, combined with the limited ICA field of view, prevents ICA from measuring the full cometary ion distribution. If too few low energy ions are measured, a distribution cannot be inferred, and so no temperature can be calculated. Thus, finding the temperature necessary for the thermal pressure requires data with a reasonably complete distribution.

Here, we select ICA data that contains a low energy cometary ion distribution representative of the surrounding plasma by selecting time periods with densities comparable to that of LAP/MIP and correct the data for the spacecraft potential. The resulting distribution is then fit to find both the drift energy and temperature of the cometary ions across a wide span of the comet's orbit, allowing us to draw conclusions on how both parameters change with heliocentric distance. This can then be used in future work to help calculate the ion thermal pressure, while also giving insight into the changing distributions of the cometary ions.

\section{Instrument description}
In this work, we use data from three instruments in the Rosetta Plasma Consortium (RPC): the Ion Composition Analyzer (ICA), the Langmuir Probe (LAP), and the Mutual Impedance Probe (MIP). The bulk of the data used is from ICA, with cross-calibrated LAP-MIP data used for ICA data selection and spacecraft potential. 

\subsection{Ion Composition Analyzer}
The Ion Composition Analyzer (ICA) \citep{Nilsson2007} was an ion mass spectrometer in the Rosetta Plasma Consortion (RPC) package \citep{Carr2007}. The instrument design consisted of an electrostatic entrance deflection system, used to determine ion arrival angle, a hemispherical electrostatic analyzer for energy analysis, a magnetic mass analyzer, and a microchannel plate detector. ICA measured ions in the energy range of a few eV/q to 40 keV/q in 96 energy steps, with $\Delta E/E = 0.07$ for $ E > 30 \mathrm{eV}$ and 0.30 for $ E < 30 \mathrm{eV}$, and had a mass resolution of $M/\Delta M \sim 2$. This mass resolution can distinguish between masses 1, 2, 4, 8, 16, and 32 amu/q. In this paper, we treat all ions with masses > 16 amu/q as water group ions originating from the comet. Each full energy-angle scan took 192 s.

ICA had a nominal angular field of view of $2 \pi$ sr, with $90^{\circ} \, \times \, 360^{\circ}$ coverage with a resolution of $5.625^{\circ} \, \times \, 22.5^{\circ}$ in elevation and azimuth respectively. However, this field of view is not applicable to the lowest energy ions, such as those discussed here. Elevation bins are measured using electrostatic deflection. Thus, the elevation range is dependent on the energy bin being measured, and so for low energies ($\lessapprox$ 100 eV), only the central elevation bins are scanned over due to difficulties in setting deflector voltages accurately. Additionally, the negative spacecraft potential highly distorts ion trajectories at these energies \citep{Bergman2020}. As such, all data used in this study was summed over all angular directions to produce a 1D distribution in energy. For the spectra we use the heavy ion data from the level 4 mass-separated Phys-Mass data \citep{Physmass}.

\subsubsection{Energy table correction}
The energy table required updating upon arrival at comet 67P, when it was discovered that the ICA high voltage system had an offset compared to on-ground calibration results \citep{Nilsson2017}. After update, the uncertainty in ion energies is roughly 3 eV. However, there is an additional high voltage drift related to the instrument temperature, which occurs when ICA cooled to below $13.5^{\circ}$C. This can be corrected with the formula:
\begin{equation}
E_c =
    \begin{cases}
        E_0 + (13.5 - T_s) \times 0.7 & \text{if } T_s < 13.5^{\circ} \text{C}\\
        E_0 & \text{otherwise}
    \end{cases}
\end{equation}
where $E_c$ is the corrected energy, $T_s$ is the sensor temperature, and $E_0$ is the original energy. This formula has been applied to all data used in this study.

\subsection{Langmuir Probe and Mutual Impedance Probe}
The Rosetta Plasma Consortium also contained a Langmuir probe instrument, LAP \citep{Eriksson2007}, and a Mutual Impedance Probe, MIP \citep{trotignon_rpc-mip_2007}. Data from these two instruments provided spacecraft potential and electron density for this study, which was compared to the ICA ion density. LAP consisted of two spherical Langmuir probes, called LAP1 and LAP2, located on two booms mounted to the spacecraft. The boom lengths were 2.2 m and 1.6 m for LAP1 and LAP2 respectively. The booms were positioned such that at least one of the probes was always sunlit and capable of measuring a radial outflow from the comet \citep{Edberg2015}. LAP primarily measured the current to the probes as a function of bias voltage, which can then be used to calculate plasma parameters such as spacecraft potential, electron density, and ion flow speed. 

The Mutual Impedance Probe had two pairs of receiving and transmitting electrodes mounted to the LAP1 boom 1 m apart. MIP then could retrieve the plasma density by finding the plasma resonance frequency. If densities were low, MIP could also use the LAP2 probe as a transmitting electrode, thus increasing the baseline to 4 m. MIP density retrievals, unlike LAP, are independent of electron temperature (as long as the plasma frequency is within its measurement range) and not sensitive to the spacecraft potential. Thus, because the two measurement techniques are so different, they have been cross-calibrated to ensure high accuracy \citep{johansson_plasma_2021}. This cross-calibrated electron density dataset \citep{eriksson_rosetta_2020} has been used in this study and is shown in Fig. \ref{fig:all_densities}. We additionally use the derived LAP ion speeds for comparison with our fit parameters. The effective ion speed (i.e., a combination of the thermal and drift speeds) can be calculated from the ion current slope of the LAP bias voltage sweep using the equation:
\begin{equation}
    \frac{dI}{dV} = \frac{q^2 n_i A_{LAP}}{2 m_i u_i}
\end{equation}
where $dI/dV$ is the slope of the ion current region, $q$ is the ion charge (assumed to be singly charged), $n_i$ the plasma density from MIP, $A_{LAP}$ is the surface area of the Langmuir probe, $m_i$ the ion mass (assumed to be 19 in the LAP dataset), and $u_i$ is the effective ion speed \citep{vigren_effective_2017}. 

\section{Methods}
\subsection {Data selection} \label{sec:data_select}
We use a few criteria to select ICA data that is both inside the cometopause and include most of the low energy ion distribution. First, we average the solar wind, cometary ion, and electron densities and pressures for every 12 hours, approximately the rotation period of the comet, to smooth out variations from asymmetric outgassing. To ensure the data is inside the cometopause, we select only data where the cometary ion momentum flux is larger than the solar wind momentum flux (calculated as in \citet{Williamson2020}). ICA ion densities are routinely lower than LAP/MIP electron densities, assumed to be due to unobserved low energy ions. Selecting ICA data where the cometary ions are a significant percentage of the electron densities should help to mitigate the issue of the missing ions. We found that setting a threshold of $n_c \geq 0.15 n_e$ where $n_c$ is the cometary ion density and $n_e$ is the LAP/MIP electron density allowed for both a significant number of time periods that meet the condition as well as ensure ICA was observing the majority of the local plasma distribution. 

\begin{figure}
 \includegraphics[width=\columnwidth]{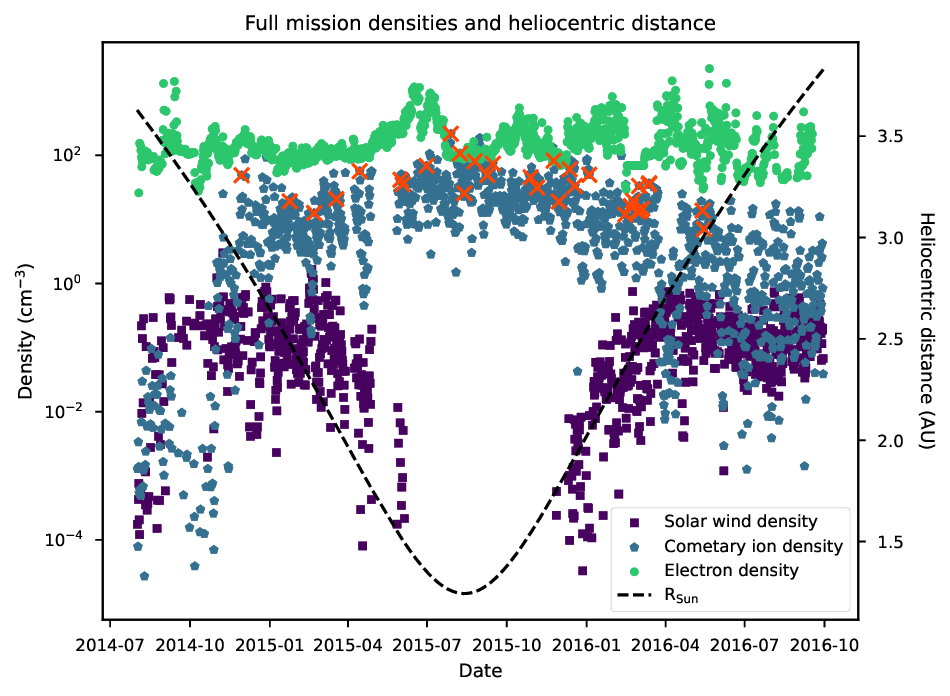}
 \caption{Electron (green), cometary ion (blue), and solar wind (purple) densities, averaged over 12 hours, for the entire Rosetta mission. Electron densities are from the LAP/MIP cross-calibrated data and cometary and solar wind ion densities are from ICA. The black dashed line shows heliocentric distance. The orange crosses mark the 32 randomly chosen time periods for fitting.}
 \label{fig:all_densities}
\end{figure}

Applying both of these conditions to the ICA data results in 259 12-hour time periods suitable for analysis, 148 of which are in the solar wind ion cavity. We then take a random sample of 32 of these time periods for use in the fitting procedure, giving 4676 suitable ICA scans to fit as ICA was not on for the entirety of every 12 hour period. Figure \ref{fig:all_densities} shows the 12-hour averaged electron, cometary ion, and solar wind densities, as well as heliocentric distance, for the entirety of the main Rosetta mission, from August 2014 to September 2016. The solar wind ion cavity is clearly visible as a gap around perihelion where the solar wind densities are non-existent, and times when Rosetta was inside the cometopause are roughly a few months on either side of the solar wind ion cavity. We show the 12-hour time periods selected for fitting with orange X markers.

\begin{figure}
 \includegraphics[width=\columnwidth]{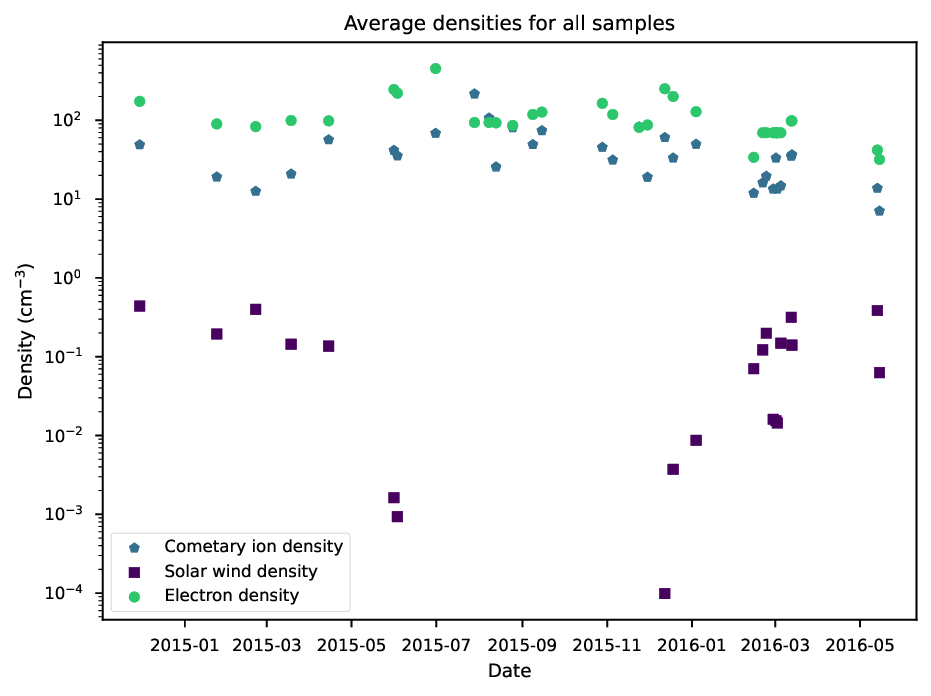}
 \caption{Electron (green), cometary ion (blue), and solar wind (purple) densities, averaged over 12 hours, for the 32 chosen time periods.}
 \label{fig:sample_densities}
\end{figure}

The densities of the 32 time periods are also shown in Figure \ref{fig:sample_densities}. We note that, even though our conditions ensure they are all inside the cometopause, the time periods still cover a wide range of conditions, with the local solar wind density varying by several orders of magnitude. Eleven of the time periods are fully inside the solar wind ion cavity, with solar wind densities equal to zero, while the other 21 have both cometary and solar wind ions present. The comparison between the cometary ion densities and electron densities, showing that they are of a similar order of magnitude, can be seen in this figure.

\subsection{Fitting method} \label{sec:fitmethod}
To determine the bulk speed (drift energy) and 1D temperature of the low energy ions, we fit a drifting Maxwell-Boltzmann distribution to the data. The same model was used previously in \citet{Bergman2021} and \citet{moeslinger_solar_2023}. The fit is done in energy space and the fit function is given by
\begin{equation} \label{eq:MB}
    f_E(E) = n \, \sqrt{\frac{1}{\pi k T_i E_i}} \, \exp \left(-\frac{E+E_i}{k T_i}\right) \, \sinh \left(\frac{2\sqrt{E E_i}}{k T_i}\right)
\end{equation}
where $n$ is the ion density, $k$ the Boltzmann constant and hence $kT_i$ is the ion temperature in eV, and $E_i = m_i v_i^2/2$ is the ion drift energy, equivalent to the bulk speed in energy space. The full derivation of this equation can be found in Appendix B of \citet{Bergman2021}. For some scans, the energy distribution has two peaks. For these cases both peaks are fit with a two-peak version of Equation \ref{eq:MB} where the total fit is a sum of the fits for each peak. As will be shown in Section \ref{sec:results}, some scan distributions have an additional higher energy tail, corresponding to pickup ions. These tails would be best fit with a kappa distribution; however, here we choose to focus only on the low energy Maxwellian peaks, as these ions comprise the bulk of the plasma. Indeed, the value of the distribution function at low energies is several orders of magnitude higher than that of higher energies, meaning the kappa distribution tail contributes little to the bulk plasma properties.

The Maxwellian model fit to the data from each scan is inspected manually and discarded if the fit quality is poor. We additionally calculate a modified index of agreement \citep{legates_evaluating_1999} for each fit to confirm the fit quality. Slightly less than 25\% of the 4676 scans could not be fit successfully, leaving 3521 scans from the 32 time periods shown in Fig. \ref{fig:sample_densities} with good quality fits.

\subsection{Spacecraft potential correction}
\begin{figure}
 \includegraphics[width=\columnwidth]{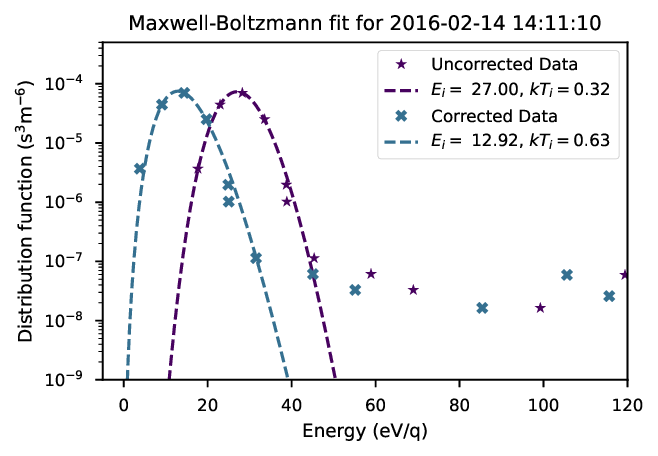}
 \caption{A single ICA scan from February 14, 2016 fit with a drifting Maxwell-Boltzmann distribution. Energy per charge in eV is shown on the x axis and distribution function ($\mathrm{s^3 m^{-6}}$) on the y axis. The original, uncorrected ICA data and corresponding fit are shown in purple, while the data corrected for a spacecraft potential of -13.7 V is shown in blue. Resulting fit parameters are shown in the legend.}
 \label{fig:cor_uncor}
\end{figure}

As described in section \ref{sec:Intro}, spacecraft potential affects ions with energies roughly $ < |2 V_{s/c}|$. As we are interested in the low energy newborn cometary ions for this work, we must correct for the acceleration due to spacecraft potential. While \citet{Bergman2020} and \citet{Bergman2020a} show that the spacecraft potential distorts the incoming direction of the low energy ions, only the one dimensional energy distribution is needed to obtain bulk speed and temperature. Thus, we only correct for the energy added to the positive ions. This is done simply by adding the average (negative) spacecraft potential to the corresponding ICA ion energies. This method was confirmed with modeling in \citet{Bergman2021}. The new ion energies per charge will then be:
\begin{equation}
    \frac{E_{new}}{q} = - V_{s/c} + \frac{E_{ICA}}{q}
\end{equation}
where $q$ is the ion charge (assumed to be one for the cometary ions). An example where a drifting Maxwell-Boltzmann distribution has been fit to the data as described in Section \ref{sec:fitmethod} is shown in Figure \ref{fig:cor_uncor}. For this sample the mean spacecraft potential was -13.7 V. The original data is shown in purple, with stars marking the measurements, while the data corrected for spacecraft potential is shown in blue, with X's marking the measurements. In the figure, it is evident that accounting for the spacecraft potential not only changes the peak energy as expected, but also changes the fitted temperature. The peak energy is always decreased and the temperature increased after the spacecraft potential correction, since the negative spacecraft potential both accelerates and adiabatically cools the ions.

\section{Results} \label{sec:results}
\subsection{Fit examples}
\begin{figure}
 \includegraphics[width=\columnwidth]{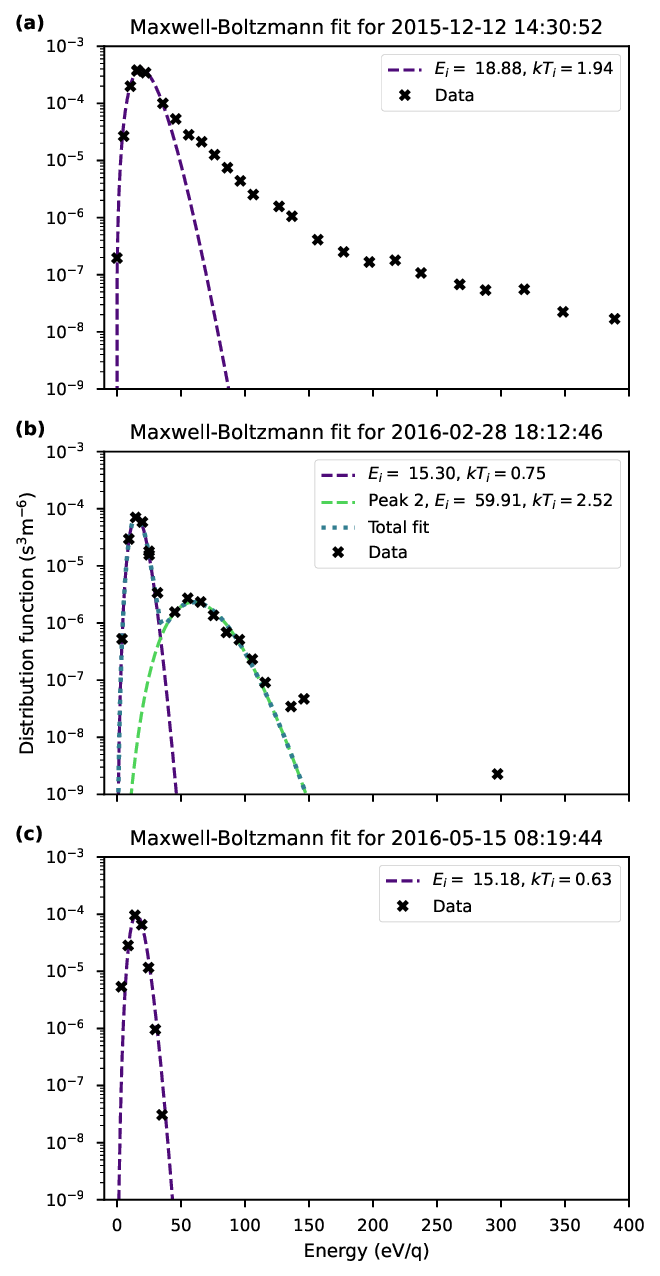}
 \caption{Examples of the three most typical energy distributions in the samples. Energy per charge in eV is shown on the x axis and distribution function ($\mathrm{s^3 m^{-6}}$) on the y axis. Data is given in black and the fits in purple. Panel (a) shows a fit with one low energy peak followed by a high energy tail, panel (b) shows a distribution with two Maxwell-Boltzmann peaks, with the additional second peak fit in green and the total fit in blue, and panel (c) shows a single low energy peak. The panels are ordered by increasing heliocentric distance.}
 \label{fig:scan_types}
\end{figure}

An example of a simple one peak fit is shown in Fig. \ref{fig:cor_uncor} and panel (c) of Fig. \ref{fig:scan_types}. A higher energy tail not fit by the drifting Maxwell-Boltzmann function can be seen in Fig. \ref{fig:cor_uncor}; however, the distribution function is typically several orders of magnitude lower than that of the data fit by the function. Many scans, particularly closer to perihelion, contain more prominent high energy tails that appear to fit an exponential or kappa distribution such as that in panel (a) of Fig. \ref{fig:scan_types}; we leave the study of these distributions to a future work. Other scans include a second clearly Maxwellian peak of smaller magnitude at higher energies. An example of this is shown in panel (b) Fig. \ref{fig:scan_types}. As these form a minority of the scans, we only include the lower energy peak in our analysis, as the second peak is always lower amplitude than the primary low energy peak, but fit the second peak when it appears to ensure the correct fit for the lower energy peak.

As mentioned previously, approximately 25\% of the scans were unable to be successfully fit. This was primarily due to two reasons: either the distribution lacked a low energy peak, or there was not enough data for fitting. Particularly close to perihelion, scans could have a broad, highly asymmetric distribution that could not be fit with the drifting Maxwell-Boltzmann, as they did not have a distinct peak. In other samples, scans occasionally lack a low energy peak entirely, indicating ICA likely did not observe the full cometary ion distribution in that scan. We also set the requirement that a scan must have at least 5 data points in order to be fit, and that the resulting parameters must have $ E_i > 5$ eV and $kT_i > 0.4$ eV as arbitrary lower limits to approximate the lowest values resolvable by the instrument based on examination of the data. The temperature lower limit was the result of the instrument energy resolution and the energy lower limit ensured there were measurements below the peak for a better fit. Scans excluded for this reason tended to be the samples farthest from perihelion, later in the mission. Thus, the ratio of skipped scans to total scans in a 12-hour sample as a function of heliocentric distance has two peaks, with one peak close to perihelion from the unsuitable distributions and the second far from perihelion due to either an unresolvable distribution or too few data points.

\subsection{Parameters from all fits}

\begin{figure}
 \includegraphics[width=\columnwidth]{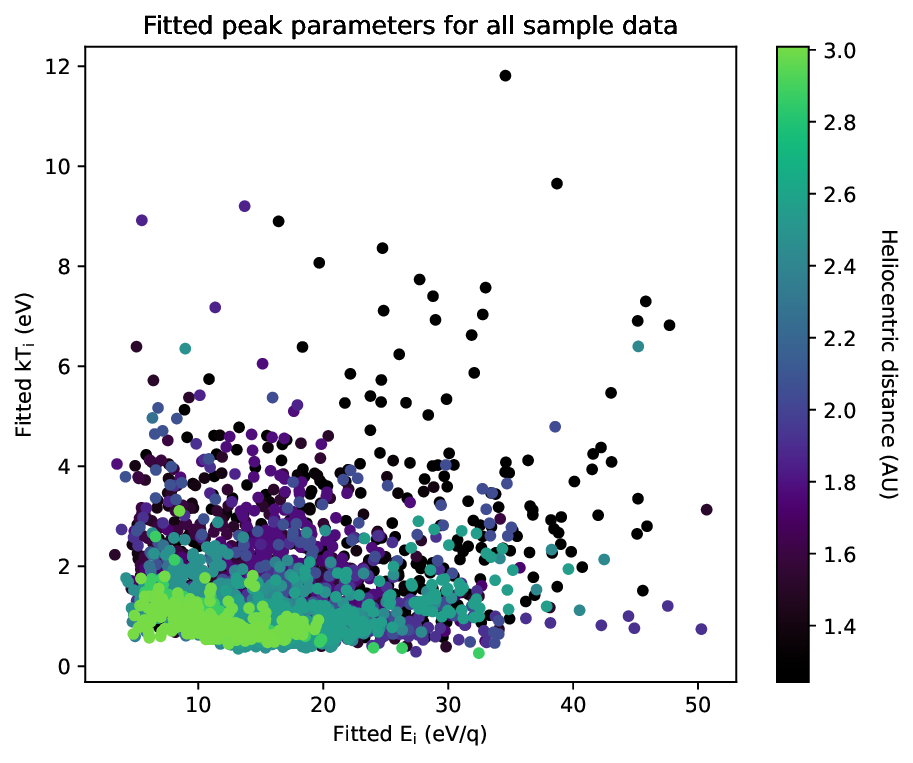}
 \caption{The peak drift energy (x axis) and temperature (y axis) for all 3521 scans with a successful fit. If a scan distribution included more than one peak, we have taken the energy and temperature of the lowest energy peak. Color indicates heliocentric distance, as shown by the colorbar on the right.}
 \label{fig:all_params}
\end{figure}

\begin{figure}
 \includegraphics[width=\columnwidth]{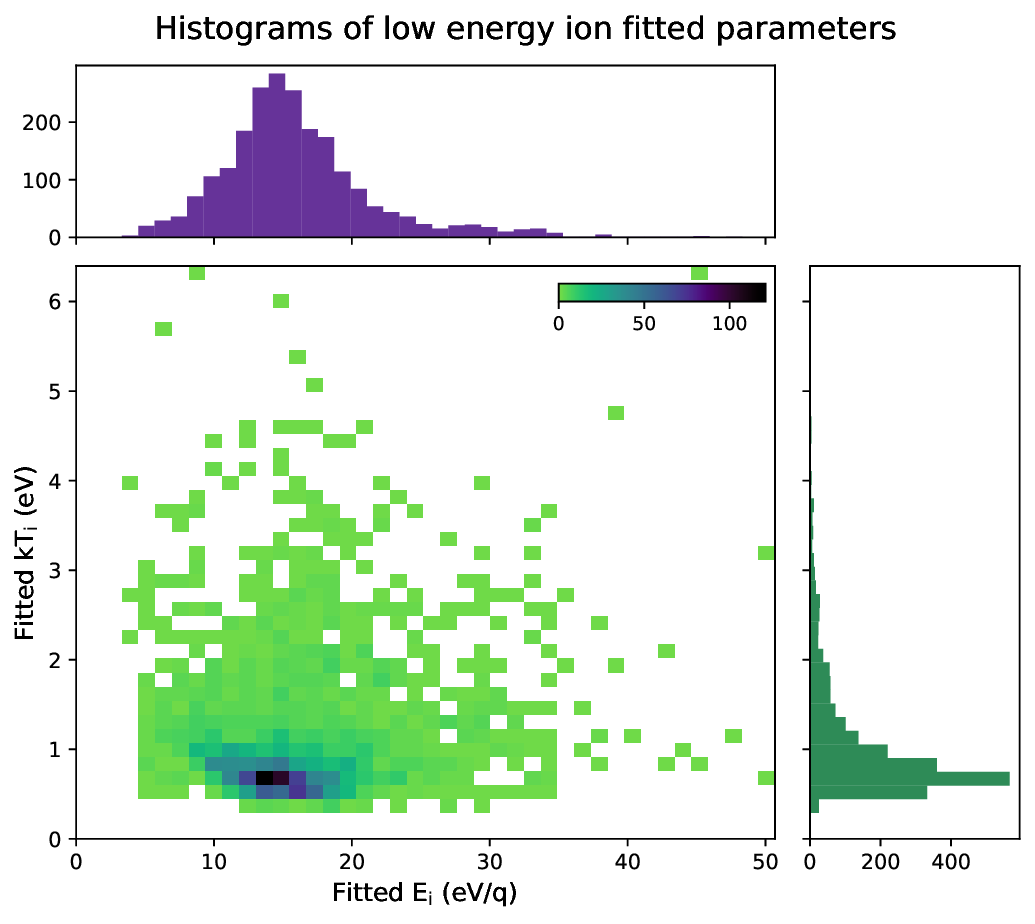}
 \caption{A histogram of all the fitted peak drift energy and temperature shown in Figure \ref{fig:all_params}. The top panel shows a histogram of the energy in purple, the middle shows a 2D histogram of both energy and temperature, and the right panel shows temperature in green.}
 \label{fig:hist}
\end{figure}

In Fig. \ref{fig:all_params} we show the resulting fit parameters, drift energy (bulk velocity) and temperature, for all successful fits, with color indicating heliocentric distance in AU. Broadly, while there is a significant amount of scatter, there is also a relatively small range of values that contain most of the data, visible in the histogram in Fig. \ref{fig:hist}. 50\% of the data has drift energies between 11 and 18 eV (approximately 11--14 km/s flow speed assuming $\mathrm{H_2O^+}$) and temperatures between 0.5 and 1 eV. Both $\mathrm{E_i}$ and $\mathrm{kT_i}$ are log-normally distributed across the whole dataset (not shown), with their individual distributions shown in the top and right panels of Fig. \ref{fig:hist}. The individual histograms for $E_i$ and $kT_i$ indicate that the lower limit cutoffs of 5 eV and 0.5 eV respectively do not drastically bias the data, as both show peaks well above the cutoff values.

\subsection{Variations with heliocentric distance} \label{sec:helio}
One area of interest is the variation of the drift energy and temperature with heliocentric distance, as this controls the size and complexity of the comet environment. Additionally, finding such a trend would allow for generalization across the broader ICA dataset. However, determining the effect of heliocentric distance is complicated by the varying spacecraft distance to the comet, which exponentially increased with decreasing heliocentric distance. The cometocentric distance in our dataset thus ranged from a maximum of 397 km at a heliocentric distance of 1.25 AU to a minimum of 9.5 km at 3 AU. The increase on spacecraft cometocentric distance near perihelion was an effect of the increasing comet activity; the spacecraft endeavored to remain in roughly the same density corridor. In theory, the changing spacecraft distance should also affect the ion distributions observed by Rosetta, and so it would be desirable to disentangle the two effects.

\begin{figure}
 \includegraphics[width=\columnwidth]{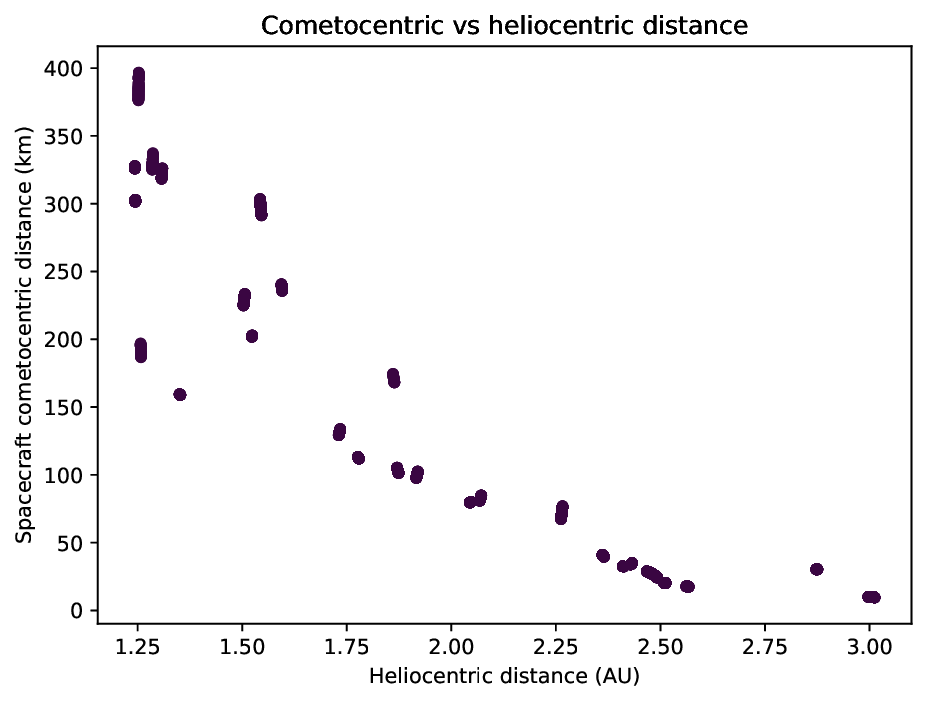}
 \caption{The distance from the comet to the Sun in AU vs the spacecraft distance to the comet in km for all the data used in this study.}
 \label{fig:rsrc}
\end{figure}

Fig. \ref{fig:rsrc} shows the strong correlation between heliocentric distance, $\mathrm{R_{sun}}$, and cometocentric distance, $\mathrm{r_{com}}$. The data examined for this study, while covering a wide portion of the Rosetta mission, simply does not cover enough parameter space where one distance varies much more than the other in order to fully disentangle the two effects. Therefore here we only show trends with respect to heliocentric distance, with discussion on the cometocentric dependence.

To ascertain trends in the data, we analyze the datasets using the Mann-Kendall test \citep{mann_nonparametric_1945,kendall_rank_1975,gilbert_statistical_1987} and Theil-Sen regression \citep{theil_rank-invariant_1992,sen_estimates_1968}. The Mann-Kendall test is a non-parametric test for monotonic trends. To assess the significance of the trend in the ICA data, we employed bootstrapping to determine the statistical distribution of the Mann-Kendall test under the null hypothesis of no trend. This involved repeated random sampling with replacement from the original dataset. The significance of the trend is indicated by the observed positive or negative deviation of the Mann-Kendall test result (of the observed data) from the null-hypothesis distribution. The test is two-tailed, meaning it is capable of determining the presence of both positive and negative trends. We evaluate the test at a 95\% confidence level. We use a Theil-Sen regression to estimate a slope for the parameter trends. Theil-Sen regression was chosen because it is also a non-parametric test, and therefore no assumptions about the distribution of the data are required. Additionally, it is less sensitive to outliers than a traditional least-squares regression, which is useful for our dataset.

From Fig. \ref{fig:all_params}, it appears that there is more scatter, and hence a larger scan-to-scan variation, in drift energy and temperature for lower heliocentric distances (darker colors). To test this assumption, we calculate the interquartile range of the variation for each sample period and validate the trend using the Mann-Kendall statistical test. The interquartile range shows a significant decreasing trend with heliocentric distance for both $\mathrm{E_i}$ and $\mathrm{kT_i}$. Thus, the plasma does indeed vary more from scan to scan at lower heliocentric distances, indicating a more dynamic plasma near perihelion.

\begin{figure}
 \includegraphics[width=\columnwidth]{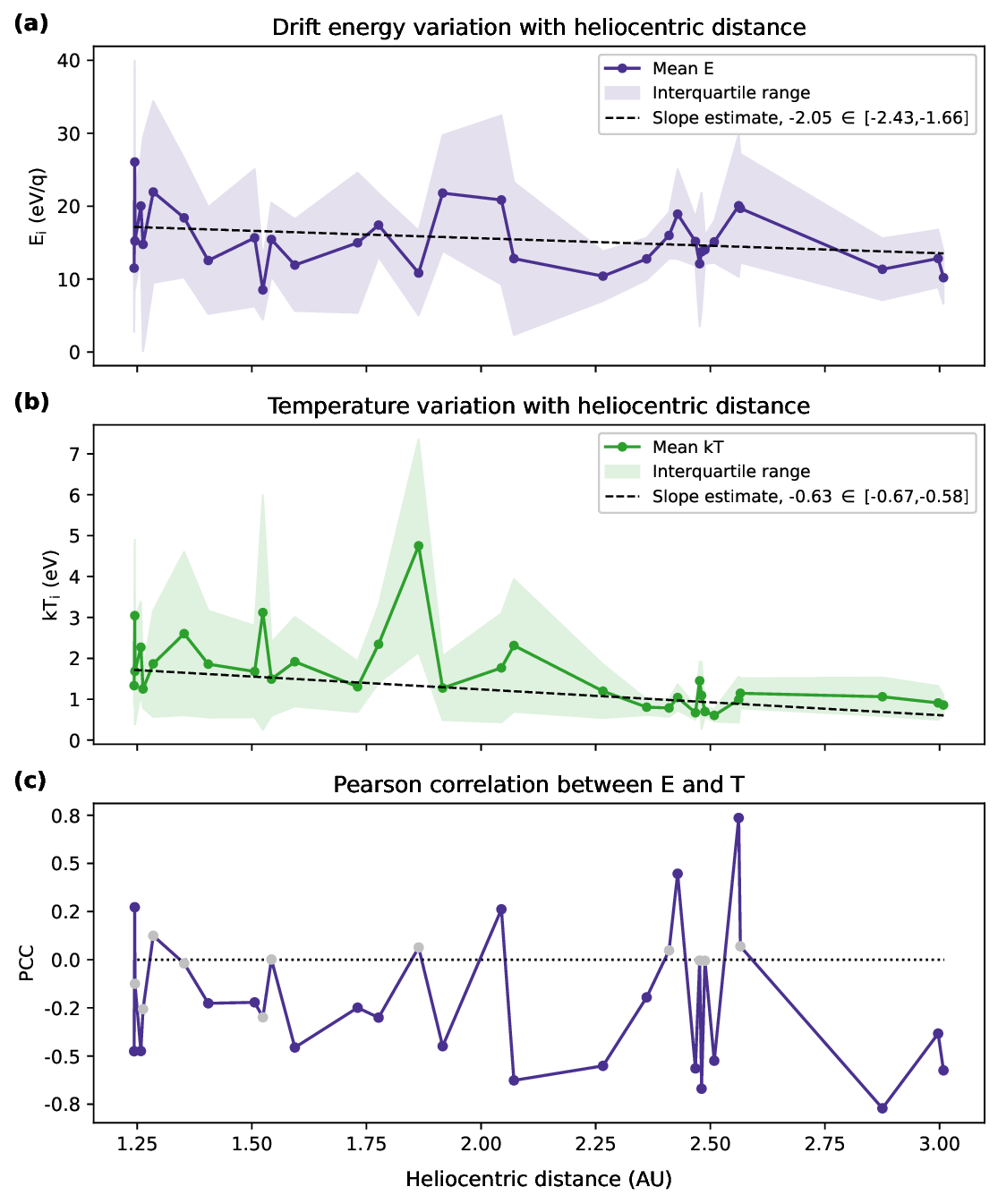}
 \caption{Variations of the fitted parameters with heliocentric distance in AU. Panel (a) shows the mean drift energy per sample (eV) in dark purple, the interquartile range (25\% - 75\%) for a sample in surrounding light blue, and the Theil-Sen slope fit as a black dashed line. Panel (b) shows the same for temperature (eV) in green. The Theil-Sen slope estimates are given in the panel legends with the 95\% confidence intervals. Panel (c) shows the Pearson correlation coefficient between E and T for each sample, with a dotted line indicating 0. Gray markers indicate a correlation coefficient with a p-value above 0.05 and are not statistically significant.}
 \label{fig:helio_means}
\end{figure}

We additionally use the Mann-Kendall test to investigate if the drift energy and temperature decrease with increasing heliocentric distance, followed by Theil-Sen regression to estimate a slope. While Fig. \ref{fig:helio_means} shows the mean parameters per 12-hour sample to minimize clutter, both the Mann-Kendall significance test and Theil-Sen estimate were done using the full dataset. Both drift energy and temperature have a highly significant decreasing trend with increasing heliocentric distance. However, the decrease is small, with the Theil-Sen regression giving trends of $-2.05 R_{sun} + 19.7$ for drift energy and $-0.63 R_{sun} + 2.5$ for temperature with $R_{sun}$ in AU. Thus, while distance from the Sun does have a statistically significant effect on the low energy ion distributions, it is not a strong one.

There is no overall strong correlation between $\mathrm{E_i}$ and $\mathrm{kT_i}$ in the full dataset, although this is not always the case for individual 12 hr samples. As seen in panel (c) of Fig. \ref{fig:helio_means}, many of the samples have a negative correlation, i.e. $\mathrm{kT_i}$ decreases as $\mathrm{E_i}$ increases. However, there are four samples with a statistically significant positive correlation, and other samples with no significant correlation at all. It is unclear why both positive and negative correlations appear, and the type of correlation does not appear to relate to heliocentric distance. Possible causes of these varying correlations will be explored in Section \ref{sec:discussion}.

Unsurprisingly, given the negative correlation between $\mathrm{R_{sun}}$ and $\mathrm{r_{com}}$, the drift energy and temperature do show a positive correlation with cometocentric distance (Fig. \ref{fig:cometo_means} in the appendix shows the cometocentric equivalent to Fig. \ref{fig:helio_means}). While the significance of this correlation is similar to those for heliocentric distance, it is highly dependent on the subset of data chosen. For example, the temperature no longer shows any trend with cometocentric distance if only data with $\mathrm{R_{sun}} < 1.6$ AU (which equates to $\mathrm{r_{com}} > 170$ km) are selected for the Theil-Sen regression. Even so, attempting to separate the effects of heliocentric vs cometocentric distance by choosing portions of data does not remove the inherent bias in the dataset, as the two distances are much more strongly correlated with each other than with either fit parameter. With the data used in this study it is therefore not possible to determine which distance has a stronger effect on the energy and temperature, although the correlation with heliocentric distance remains even if a subset of the data is chosen. The relative importance of heliocentric vs cometocentric distance in the trends in energy and temperature are discussed below.

\section{Discussion} \label{sec:discussion}

As stated previously, all data used in this study was selected based on the average cometary ion density being at least 15\% that of the LAP electron density. However, while this is true for the 12 hr averages, it does vary on a scan-to-scan basis and helps to illuminate one possible reason ICA densities are generally lower than LAP densities. In some ICA scans, the low energy Maxwellian disappears and the distribution function overall has much lower amplitude. As the density is the integral of the distribution function, this subsequently drops as well. Thus, during times of comparatively low densities, it would seem ICA is not detecting the low energy cometary ions. Likely this is at least partially due to a field-of-view effect, as the spacecraft potential severely distorts the trajectories of the low energy ions. Near perihelion counterstreaming cold cometary ions were observed in the region of the diamagnetic cavity \citep{Bergman2021a}, indicating it is indeed possible for multiple cold ion beams to exist, which may not all be in the ICA field-of-view. Because of the trajectory distortion from the spacecraft potential, the field-of-view is increased, although directional information is no longer accurate. It is therefore possible that scans with a relatively complete energy distribution include ions that would normally not be detected by the instrument, allowing us to make a better assessment of the plasma properties. Comparison of the fitted drift energies with the LAP ion speeds shows generally good agreement to within 1--2 sigma of each other, indicating ICA is likely measuring the majority of the plasma. Some examples of the LAP ion speeds compared to those found here can be seen in Appendix \ref{appx:lap}. However, it should be noted that despite the consistency of the ICA and LAP-MIP data, drift energies in this study are generally higher than previously reported LAP values of 2--8 km/s, approximately 0.4--6 eV for $\mathrm{H_2O^+}$ \citep[e.g.,][]{vigren_effective_2017}. It is possible that ICA can only detect the full ion distribution when flow speeds are relatively high, which could explain the difference between the bulk speeds found here and in previous studies. Despite this caveat, the LAP-MIP comparison indicates that the ion distribution data used in this work are reasonably complete and useful for scientific interpretation. 

As shown in section \ref{sec:helio}, while we show trends with heliocentric distance, it is likely that cometocentric distance is also a contributor to the changes in drift energy and temperature. Certainly, a larger cometocentric distance can contribute to the increase in E and T close to perihelion for several reasons. For example, at farther cometocentric distances, the same instrument field of view covers a wider spatial area with significant ion production. Thus, ICA will observe ions produced at different distances that have experienced differing amounts of acceleration. Such observations of ions originating from different distances can increase the measured temperature. In general, the longer ion travel time from the ionization point to the spacecraft allows for more processes to affect the plasma, many of which will increase either the drift energy or temperature. However, some of these processes are also correlated with cometary activity, as the comet ionosphere is larger and more fully developed close to perihelion. Here, we focus on electric fields and wave activity as drivers of the increase in drift energy and temperature. 

As explained in the introduction, the comet environment is largely governed by three different electric fields, in order of increasing distance from the comet: the ambipolar field, the polarisation electric field, and finally the solar wind convective electric field. The ambipolar field, with its magnitude of $\sim 1/r_{com}$, may be one reason the mean drift energy is higher closer to perihelion when both the comet atmosphere size and spacecraft distance from the nucleus increase. However, effects from all three of these fields can be observed in the ion energy spectra. Although flow direction is not shown here, the additional higher energy ion populations seen in panels (a) and (b) of Fig. \ref{fig:scan_types} have a strong antisunward velocity component, indicating acceleration by the polarisation or solar wind electric fields. The second Maxwellian peak is likely ions born farther from the spacecraft that have been accelerated by the polarisation electric field or begun the pickup process. Likewise, the high energy ions have been seen as a signature of pickup ion detection \citep{Behar2016,Bercic2018,Broiles2015,Coates2015,moeslinger_indirect_2023,Nicolaou2017,williamson_development_2022}, with the highest energy ions likely coming from far upstream of the spacecraft. This mixed population is thus no longer in thermal equilibrium and so the distribution evolves from the drifting Maxwell-Boltzmann to a more kappa-like distribution. Thus, from the bottom to the top of Fig. \ref{fig:scan_types} (i.e., decreasing heliocentric distance), we see the evolution of the ion spectra away from equilibrium, although the cold Maxwellian population remains.

Waves also cannot be ignored as a cause for the inverse correlation with heliocentric distance, particularly for temperature. Indeed, the presence of a Maxwell-Boltzmann distribution itself is indication of wave activity, as the ions are born with the temperature of the neutral gas, with a distribution function that then evolves towards a Maxwellian as it expands (see e.g. Fig. 5 in \citet{vigren_1d_2017}). Many types of plasma waves were observed at comet 67P by Rosetta, including lower hybrid waves at low cometocentric distances \citep{Andre2017,Karlsson2017,StenbergWieser2017}, singing comet waves generated by pickup ions \citep{Goetz2020a}, ion acoustic waves \citep{Gunell2017,Gunell2017a}, possible ion Bernstein waves \citep{Odelstad2020}, and mirror mode waves \citep{tello_fallau_revisiting_2023,Volwerk2016}. With a larger comet ionosphere and longer travel distance to the spacecraft closer to perihelion, there is a higher likelihood that the ions measured by ICA have experienced wave heating. Thus, wave activity is a likely explanation for higher temperatures at lower heliocentric distances. Because wave activity is highly transient, it is also a logical reason for the large scan-to-scan variability in both energy and temperature. This is the case throughout all the data, as seen in Fig. \ref{fig:helio_means}, but is more prominent closer to perihelion. In general, the comet ionosphere has the highest energy input close to perihelion, and subsequently the largest and most complex structure. This then is likely the reason for the larger interquartile range of both E and T close to perihelion.

It can be seen in panel (c) of Fig. \ref{fig:helio_means} that, while the majority of the samples have a negative correlation between E and T, this is not consistent, and shows no trend with heliocentric distance. At times, indicated by the green dots in the figure, there is no statistically significant correlation at all. As with the high variability, this is likely due to the transient and dynamic nature of the comet ionosphere. When examining E vs T for individual samples, some cases show a clear shift in time, moving from e.g. a cluster of scans with relatively low E and high T to later scans with higher E and lower T. Such a change is a clear indication of an event occurring, although whether this is due to a change in the upstream solar wind which then affects the ionosphere or due to a more localized event is not possible to determine from the data. One such shift in time occurs for the most positively correlated sample seen in Fig. \ref{fig:helio_means}(c), with E and T both being originally low, then suddenly increasing, where they remain high for the rest of the time period. The cluster of data points from the beginning of the sample is only weakly positively correlated, but both a stronger correlation in the second cluster of data points and the shift to higher E and T create a strong positive correlation for the whole sample. Thus, this shifting in time can affect the correlation for a single sample, meaning that the changing plasma conditions also affect the correlation between E and T. 

\section{Conclusions}
We fit more than 3500 energy scans from the ICA instrument with a drifting Maxwell-Boltzmann function. By correcting for spacecraft potential and only choosing time periods with densities comparable to other instruments, we are able to focus on the low energy cometary ions that would otherwise be difficult to analyze. By doing so, we see that the drift energy and temperature are generally within a relatively narrow range of 11--18 eV and 0.5--1 eV, respectively, for much of the Rosetta mission. This is despite large changes in the comet ionosphere during the time periods analyzed, such as the formation of the solar wind ion cavity. The drift energy (equivalent to flow speed) is higher than the mission-wide medians reported from the Rosetta LAP data \citep{johansson_plasma_2021}; this may be due to selection bias, as the flow may need to be higher energy in order to be observed in full by ICA. However, for the scans analyzed, the fitted drift energy is consistent with the coinciding LAP data. 

While the drift energy and temperature do not change drastically throughout our samples, we do see two statistically significant changes: both energy and temperature decrease with increasing heliocentric distance, as does the interquartile range or scan-to-scan variability. This can largely be attributed to the comet ionosphere being both larger in size and more complex at the lowest heliocentric distance, although whether the correlations are strictly due to heliocentric distance or spacecraft distance from the comet, which increased towards perihelion, cannot be determined. We attribute the higher energy and temperature to both the accelerating electric fields and wave-particle interactions, which have more opportunities to accelerate or heat the ions at low heliocentric/large spacecraft distances. As the comet ionosphere increases in complexity, we can observe an evolution of the cometary ion distribution from a single cold Maxwellian population to include higher energy accelerated pickup ions.

\section*{Acknowledgements}
The work of H. Williamson and A. Johansson was supported by Swedish National Space Agency grant 2021-000105. A. Moeslinger was supported by SNSA grant 132/19. The authors thank the members of the Rosetta Plasma Consortium team for their helpful discussions.

\section*{Data Availability}
The Rosetta RPC-ICA, RPC-LAP, and RPC-MIP data are publicly available through the ESA Planetary Science Archives at \url{https://psa.esa.int/}. Fit parameters for all samples used in the paper are available at \url{https://data.irf.se/data/williamson2024mnras/}. Mann-Kendall tests were performed with the \href{https://pypi.org/project/pymannkendall/}{pyMannKendall} software package \citep{hussain_pymannkendall_2019}. Scipy v1.12.0 was used for Theil-Sen regression \citep{2020SciPy-NMeth}. Colormaps were used courtesy of \href{https://cmasher.readthedocs.io/}{CMasher} \citep{CMR_colors}.



\bibliographystyle{mnras}
\bibliography{references} 

\appendix

\section{Variations with cometocentric distance}

\begin{figure}
 \includegraphics[width=\columnwidth]{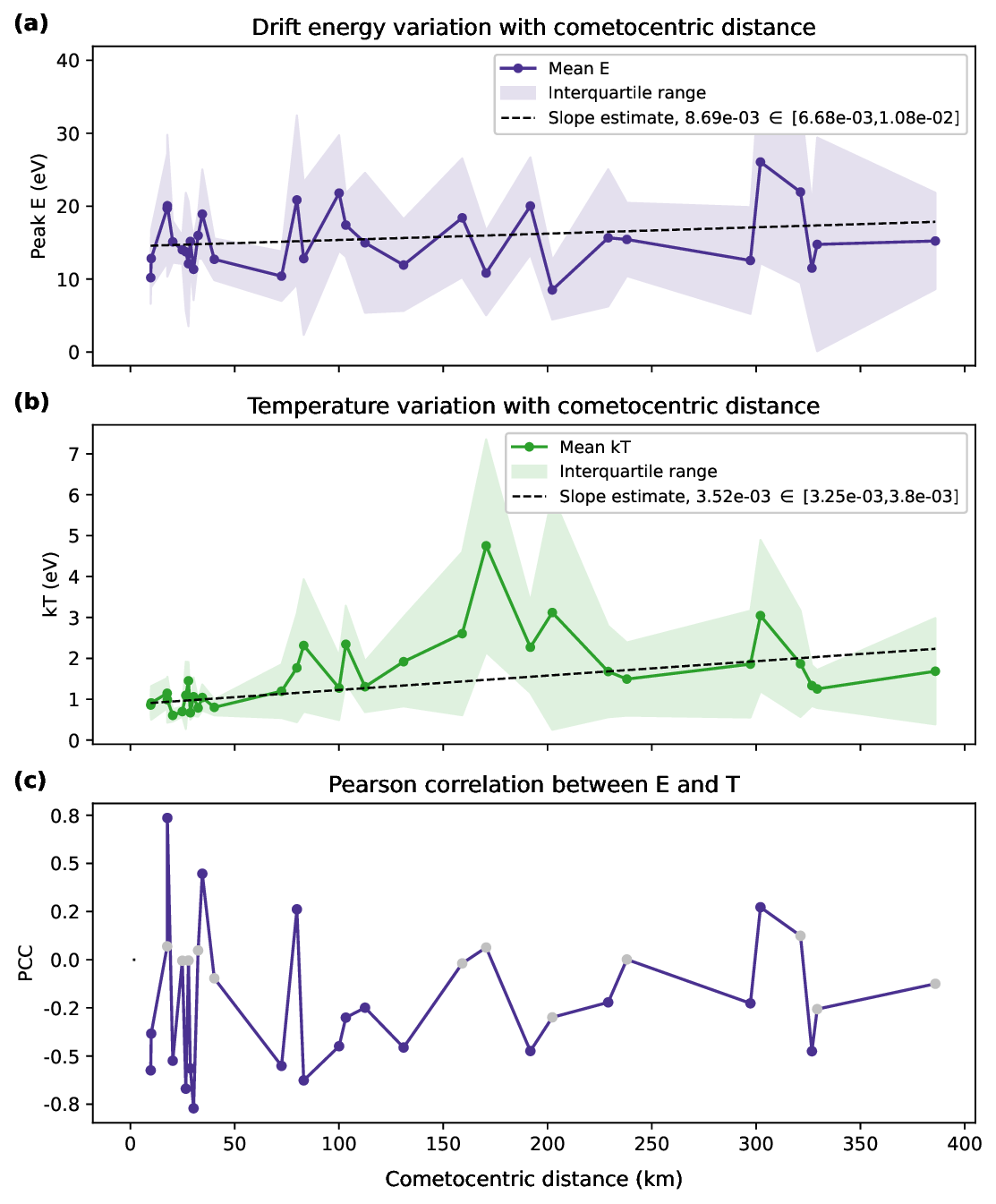}
 \caption{Variations of the fitted parameters with cometocentric distance in km. Panel (a) shows the mean drift energy per sample (eV) in dark purple, the interquartile range (25\% - 75\%) for a sample in surrounding light blue, and the Theil-Sen slope fit as a black dashed line. Panel (b) shows the same for temperature (eV) in green. The Theil-Sen slope estimates are given in the panel legends with the 95\% confidence intervals. Panel (c) shows the Pearson correlation coefficient between E and T for each sample, with a dotted line indicating 0. Gray markers indicate $ p > 0.05$ and are not statistically significant.}
 \label{fig:cometo_means}
\end{figure}

Figure \ref{fig:cometo_means} shows the fit parameters as a function of cometocentric distance in the same format as Fig. \ref{fig:helio_means}. As with the previous figure, only the mean values for each sample are plotted, but the Theil-Sen slope estimate and the Mann-Kendall trend and significance were calculated using the full dataset. Both energy and temperature have a statistically significant increasing trend with cometocentric distance.

\section{LAP speed vs ICA fitted speed} \label{appx:lap}
\begin{figure}
 \includegraphics[width=\columnwidth]{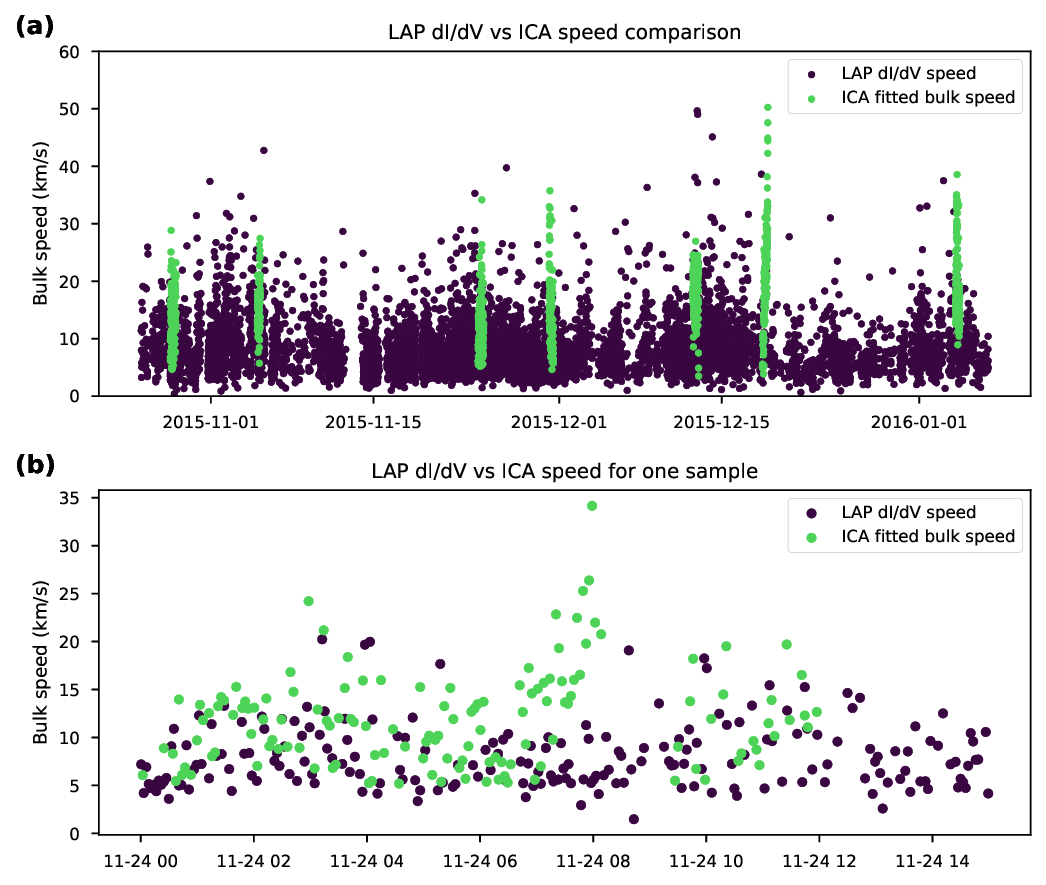}
 \caption{The LAP dI/dV effective ion flow speed in purple compared to the results from this study in green. Panel (a) shows data for two months that contain multiple fitted sample periods, while panel (b) shows a zoom in on one of the samples, November 24, 2015. The ICA fitted drift energy has been converted to bulk speed in km/s.}
 \label{fig:lap_vs_ica}
\end{figure}

We show some examples of the LAP effective ion flow speeds, derived from $dI/dV$ for each bias voltage sweep, compared to the $E_i$ from the fitted ICA data. The ICA fitted drift energy has been converted to bulk speed assuming a particle mass of 19 amu, consistent with the LAP dataset. The LAP effective ion flow speed data was taken from the analyzed sweep (ASW) data in \citet{eriksson_rosetta_2020}, filtered for quality values (a measure that depends on the goodness of the fit on the ion current slope and the MIP spectrum quality) above 0.7. The figure shows that the LAP and ICA derived speeds generally overlap, although they are not typically the exact same. In many individual samples, such as that in Fig. \ref{fig:lap_vs_ica}(b), they also show similar trends in time.

\bsp	
\label{lastpage}
\end{document}